\documentclass[
  reprint,
  superscriptaddress,
  amsmath,amssymb,
  aps,
  prab,
]{revtex4-2}

\usepackage{graphicx}   % figures
\usepackage{dcolumn}    % decimal alignment in tables
\usepackage{bm}         % bold math
\usepackage{siunitx}    % units
\usepackage{hyperref}   % hyperlinks
\usepackage{microtype}  % better spacing
\usepackage{xcolor}
\usepackage{upgreek}
\usepackage{soul}

\graphicspath{{figures/}}
\DeclareSIUnit\clight{\text{\ensuremath{c}}}
\DeclareSIUnit[per-mode=symbol]\GeVoverc{\GeV\per\clight}

\begin{document}

\title{Fast optics-based modeling enabling large-scale optimization of the H4 and M2 beamlines in the CERN SPS North Area}

\author{Giovanni Dal Maso}
\email{giovanni.dal.maso@cern.ch}
\affiliation{CERN, BE-EA, European Organization for Nuclear Research, CH-1211 Geneva, Switzerland}
\author{Dipanwita Banerjee}
\affiliation{CERN, BE-EA, European Organization for Nuclear Research, CH-1211 Geneva, Switzerland}
\author{Nikolaos Charitonidis}
\affiliation{CERN, BE-EA, European Organization for Nuclear Research, CH-1211 Geneva, Switzerland}

\begin{abstract}
The CERN multi-purpose secondary beamlines are invaluable facilities offering a rich and diverse physics
program to the international particle physics community, providing many different particle species and
beam characteristics to fixed-target experiments and test-beam users. A number of magnetic elements are
required to enable such flexibility, posing a significant challenge in the design and optimization
of the beamline optics.

In this work, we present the optimization of the H4 and M2 beamlines of the CERN
SPS North Area using genetic algorithms.
The aim of this study is to demonstrate the effectiveness of fast optics-based tracking models (here
Xsuite), to enable large-scale, multi-objective optimization on standard computing hardware. The
model is benchmarked and validated with high-fidelity Monte Carlo simulations using \textsc{BDSIM} and
cross-validation of the results with experimental measurements.

The optimized optics, applied to the H4 electron configuration for NA64, achieved a \SI{30}{\percent}
increase in available electron rate and a five-fold reduction in beam-related background. In the M2 line,
the optimized configurations yielded a nearly two-fold increase in electron rate and a \SI{67}{\percent}
improvement in muon transmission.

\end{abstract}

\maketitle

% --------------- Main content -------------
\section{Introduction}
\begin{figure}
    \includegraphics[width=\linewidth]{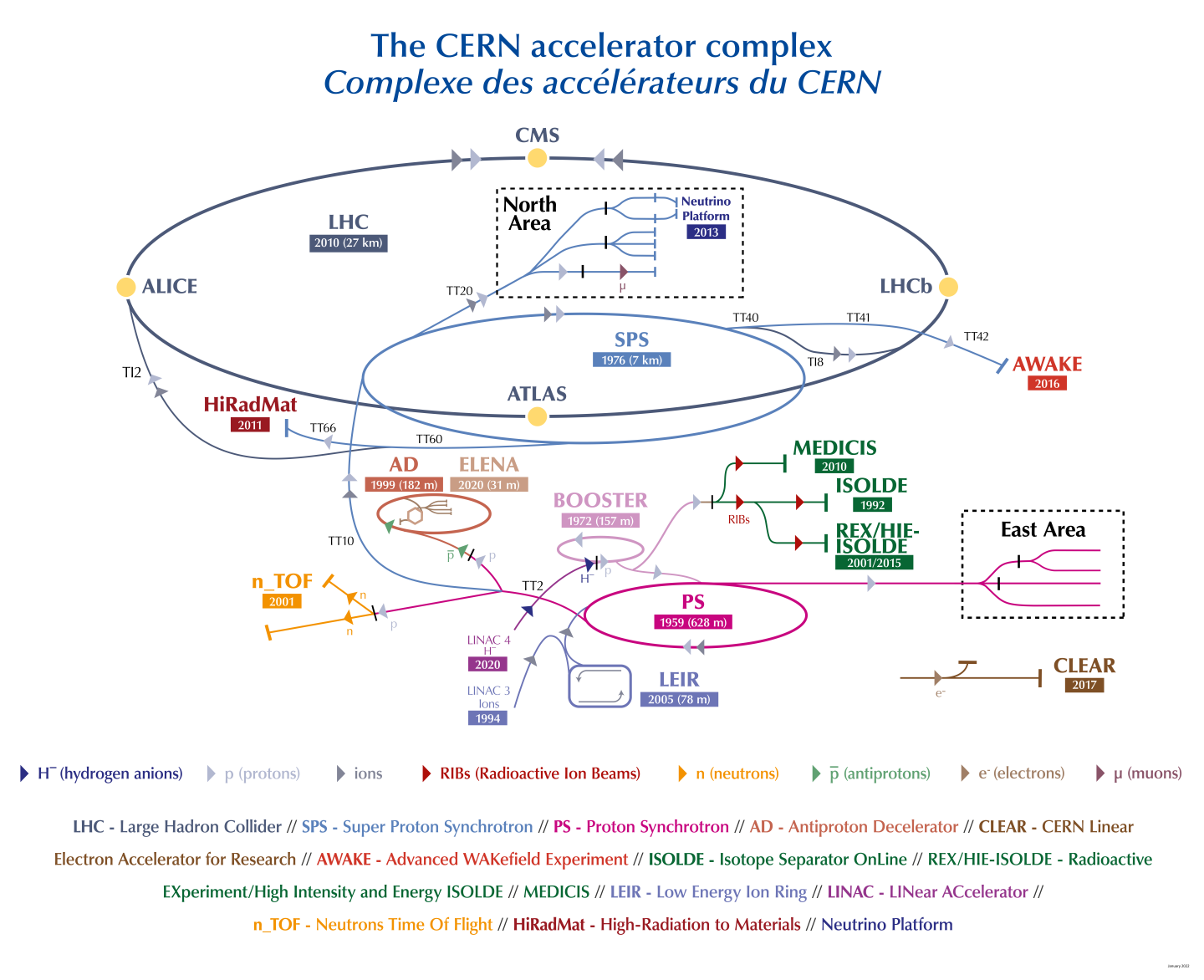}
    \caption{
        Schematic of the CERN accelerator complex.
    }
    \label{fig:CERNcomplex}
\end{figure}

The North Area Experimental Halls at CERN (EHN1, EHN2, and ECN3, see Fig.~\ref{fig:CERNcomplex}) are multipurpose facilities
hosting a broad range of fixed-target experiments and detector R\&D test-beam programs. They
are served by a network of secondary beamlines fed by three target stations, delivering a wide
spectrum of particle species, hadrons, leptons, and ions, over a momentum range from
\SIrange{10}{400}{\GeVoverc}. The secondary beams are produced by the interaction of
the \SI{400}{\GeVoverc} proton beam slowly extracted from the Super Proton Synchrotron
(SPS) on thin beryllium plate targets located approximately \SI{15}{\meter} underground. In
addition to secondary beams, primary proton and attenuated ion beams can also be delivered
directly to the experimental areas. A detailed description of the North Area facilities and their
design principles can be found in Ref.~\cite{Banerjee:2774716}.

The number of experiments and test-beam users hosted in the CERN North Area has steadily
increased in recent years, with yearly over 200 user teams for detector R\&D and fixed target experiments. Apart from the already approved experiments, new physics programs within the Physics Beyond
Colliders framework (PBC)~\cite{PBC,AlemanyFernandez:2927631} are being studied, developed and planned. 
For several of these programs, higher beam rates are a requirement for their sensitivity goals, the MUonE experiment~\cite{MUonE:LOI} being one such example, while for others they would allow the required statistics to be reached in less time. Delivering higher beam rates therefore provides a direct means to increase the scientific output within the available beam time.

To accommodate the diverse experimental requirements, the North Area beamlines are equipped
with a large number of magnetic optical elements, providing significant operational flexibility
but also leading to a high-dimensional parameter space. This makes systematic optimization of the
beam optics non-trivial. In this context, large-scale optimization techniques~\cite{9627116,9627138},
such as genetic algorithms, offer an effective framework to address multiple figures of merit and
operational constraints simultaneously. Their application to real beamlines, however, typically
requires thousands of evaluations, making the use of fast simulation tools essential in practice.

\begin{figure*}
    \includegraphics[width=\linewidth]{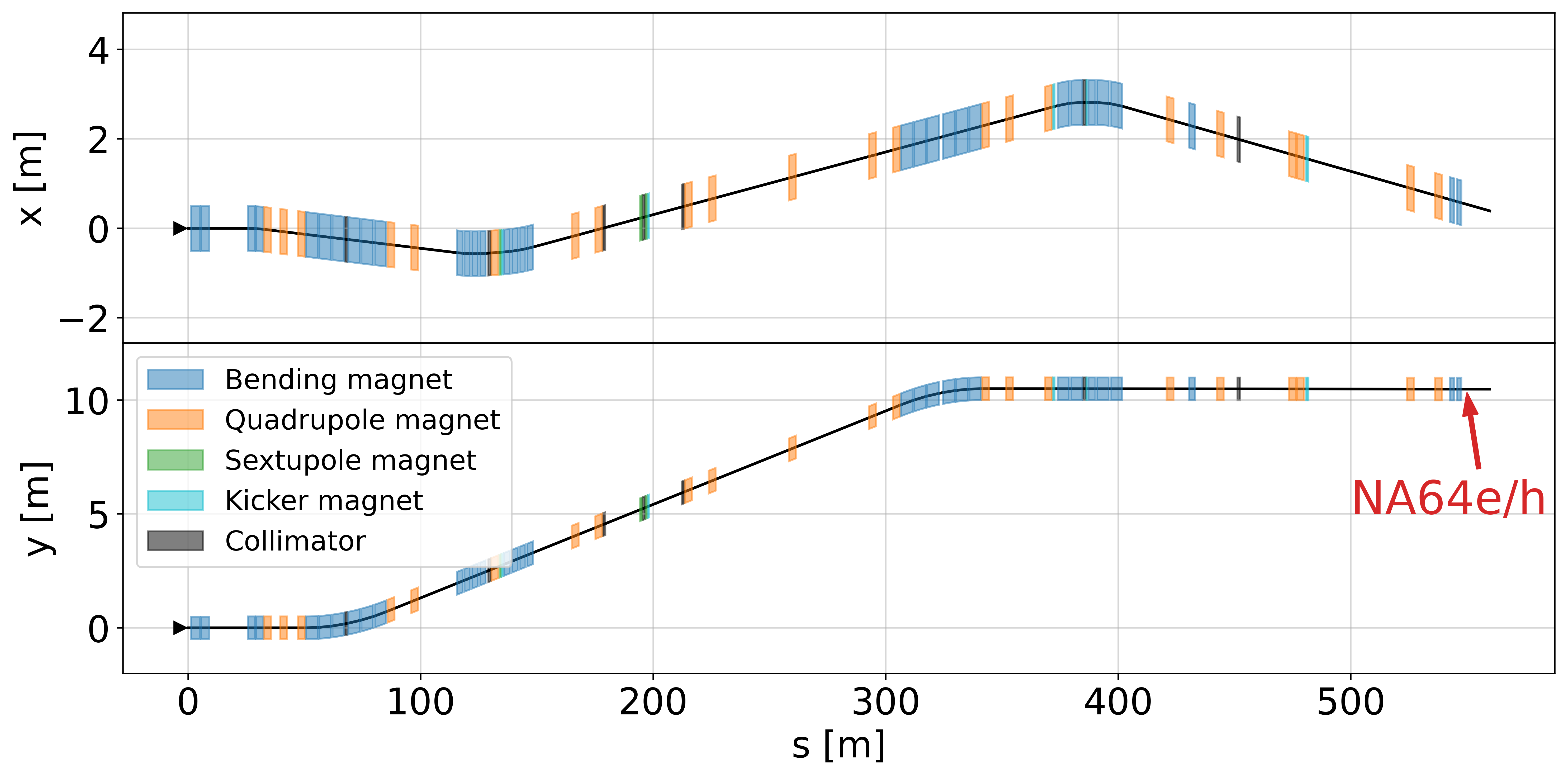}
    \caption{
        Schematic layout of the H4 beamline starting from the T2 target station.
        Dipole magnets are shown in blue, quadrupoles in orange, sextupoles in green,
        and kickers in light blue. Collimators are indicated by black markers.
        The position of the NA64 experiment is indicated.
    }
    \label{fig:h4layout}
\end{figure*}

In this work, we present a general optimization workflow and its application to two representative
beamlines of the CERN North Area: H4 (see Fig.~\ref{fig:h4layout}) and M2 (see Fig.~\ref{fig:m2layout}).
The H4 beamline, hosted in Experimental Hall North 1 (EHN1), serves various detector R\&D test-beams and the NA64 experiment, which searches for light dark matter and
dark sector mediators using electron~\cite{Andreev:2023}, positron and hadron beams~\cite{Andreev:NA64h2024}.
The M2 beamline, hosted in Experimental Hall North 2 (EHN2), delivers the highest-intensity high-energy muon beam in the world, together with high-energy hadron beams, and
currently hosts the NA64$\upmu$ program~\cite{Andreev:2024mu}, the MUonE experiment~\cite{MUonE:LOI},
and the AMBER experiment~\cite{Adams:2676885}.

We show the effectiveness of using fast optics-based tracking models, here implemented in Xsuite~\cite{Xsuite}, to enable large-scale, multi-objective optimization on standard computing hardware. The models are benchmarked and validated with high-fidelity Monte Carlo simulations using \textsc{BDSIM}~\cite{NEVAY2020107200}\footnote{\textsc{BDSIM} version 1.7.7, built on Geant4~\cite{ALLISON2016186,1610988,AGOSTINELLI2003250} 10.7.2.3 with the \texttt{FTFP\_BERT} physics list}, featuring detailed 3D geometries implemented with \texttt{pyg4ometry}~\cite{WALKER2022108228}.
The predicted improvements are compared, and found to be in good agreement, with the measurements performed during the 2025 run. The electron rate delivered to the NA64 experiment was increased by \SI{30}{\percent} with a five-fold reduction in the beam-related background, thanks to a lower beam halo content. The electron rate delivered by the M2 beamline was increased by nearly a factor two and the new optics were deployed for the calibration of the electromagnetic calorimeter of the MUonE experiment. Finally, the optimized muon optics for M2 yielded an improvement in rate by two-thirds.

\begin{figure*}
    \includegraphics[width=\linewidth]{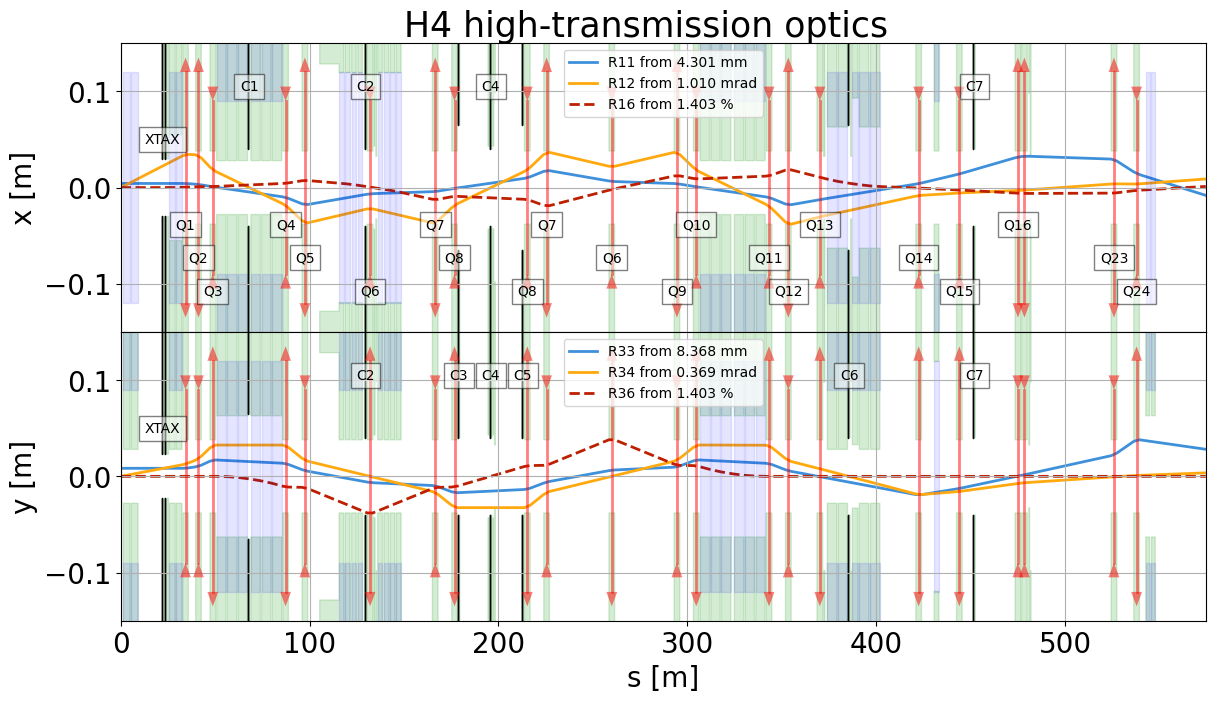}
    \caption{
        High-transmission optics of the H4 beamline starting from the T2 target center.
        The R-matrix transport elements \cite{Brown:1968hkv} are shown for the horizontal (top) and vertical (bottom) planes, scaled
        to the maximum acceptance of the beamline.
        Quadrupoles (focusing/defocusing) are indicated by outward/inward double-headed red arrows,
        dipoles by blue boxes, apertures in green, and collimators as black lines.
        Quadrupoles with identical labels are powered in series by the same power converter.
        The labels of the collimators are shown only in the plane they act on.
    }
    \label{fig:h4optics}
\end{figure*}

\section{Beamlines overview}
\subsection{The H4 beamline}

\subsubsection{Layout and optical design}
The H4 beamline is a high-energy, high-resolution secondary beamline served by the T2 target
station in the CERN North Area. A schematic overview of the beamline layout is shown in
Fig.~\ref{fig:h4layout}. The beam composition can be controlled by inserting material at dedicated locations along the beamline.
The electron beam is produced
from the \SI{400}{\GeVoverc} primary proton beam impinging on the T2 beryllium
target \cite{NA642023168776}: among the secondary particles, neutral pions decay promptly into photons, and the
resulting neutral beam is collimated by a movable collimator (XTAX) with selectable aperture shapes controlling
the transmitted beam intensity. When operating the electron beam mode, a \SI{4}{\milli\meter} thick lead converter
is subsequently used to produce electron--positron pairs by conversion.
During operation, the strengths of the H4 elements are scaled to account for the energy loss induced by synchrotron radiation. In the study presented here this effect is neglected.

The central momentum selection is provided by two vertical bends, each consisting of 6 dipole magnets, with identical bending angle
and opposite polarity. The vertical dispersion is recombined by two quadrupoles
placed symmetrically between the bends. They are respectively placed at \SI{90}{\degree} and \SI{270}{\degree} phase advance of the dispersive
array, such that their effect on dispersion is maximal.
Q6, located approximately \SI{130}{\meter} downstream of T2, is one such element.
Three horizontal bends, overall consisting of 12 dipole magnets, are included to spatially separate
H4 from the neighboring H2 beamline, which is served by the same target station. The horizontal
dispersion can be recombined by the same means; in practice, a small residual horizontal
dispersion is tolerated to allow finer control of the delivered rate. The H4 lattice includes
22 quadrupole magnets up to the NA64 experimental area, wired to 18 independent power converters,
such that the optics of the beamline is fully defined by 18 free parameters.

\subsubsection{Nominal operating modes}
The H4 beamline is currently operated in three main optical configurations \cite{Coet:172394}, selected exclusively
through changes in the quadrupole settings: high-transmission, high-resolution, and filter mode.
The \textit{three modes} were developed to cover different experimental needs and all three are regularly
used today depending on the experiment being served.

The high-transmission mode, shown in Fig.~\ref{fig:h4optics}, is the nominal configuration for
the NA64 experiment~\cite{NA642023168776}. It provides the largest angular acceptance for
particles produced in the T2 target and, combined with the maximum momentum acceptance of
$\pm\SI{1.4}{\percent}$, delivers the highest possible transmission rate at the NA64 experiment
position. The high-resolution mode trades off transmission in favor of improved momentum
resolution, while the filter mode is optimized to produce beam foci at the location of
additional absorber material, which can be inserted to enhance beam purity and is typically
used for detector R\&D.

A total of eight collimators (seven up to the NA64 experiment) are distributed along the
beamline to control the delivered rate, reduce halo, and define the momentum acceptance.
In the high-transmission optics, the momentum selection collimator C2 (see Fig.~\ref{fig:h4optics}), located approximately
\SI{130}{\meter} downstream of T2, is positioned upstream of Q6 at a point of tight vertical
focus and maximum dispersion. This condition, while necessary for sub-percent momentum
selection, represents a strong constraint in the optics design. For the NA64 experiment,
the electromagnetic calorimeter provides an energy resolution of 3--4\%, such that beam
momentum spreads of the order \SI{2}{\percent} are acceptable; furthermore, since the experiment
reconstructs the momentum per particle with a magnetic spectrometer, a modest increase in
momentum spread does not degrade the physics performance. The optimization presented in this
work therefore relaxes this constraint, no longer requiring the optics to produce a point of maximum dispersion and vertical focus at C2, in favor of increased momentum acceptance.

\subsubsection{Simulation input}
In this study, we focus on the \SI{100}{\GeVoverc} electron beam delivered to the
NA64 experiment~\cite{NA642023168776}. The input particle distributions for both the Xsuite
and \textsc{BDSIM} models of the H4 beamline were generated using a \textsc{BDSIM} simulation of the T2 target
area with \SI{e8}{} protons on target (PoT), selecting the largest XTAX aperture. To reduce the
variance of the Monte-Carlo simulation, the resulting photon distribution was resampled on the lead
converter 100 times, corresponding to an electron sample equivalent to that produced by \SI{e10}{PoT}.

\subsection{The M2 beamline}
\subsubsection{Layout and beam characteristics}
\begin{figure*}
    \includegraphics[width=\linewidth]{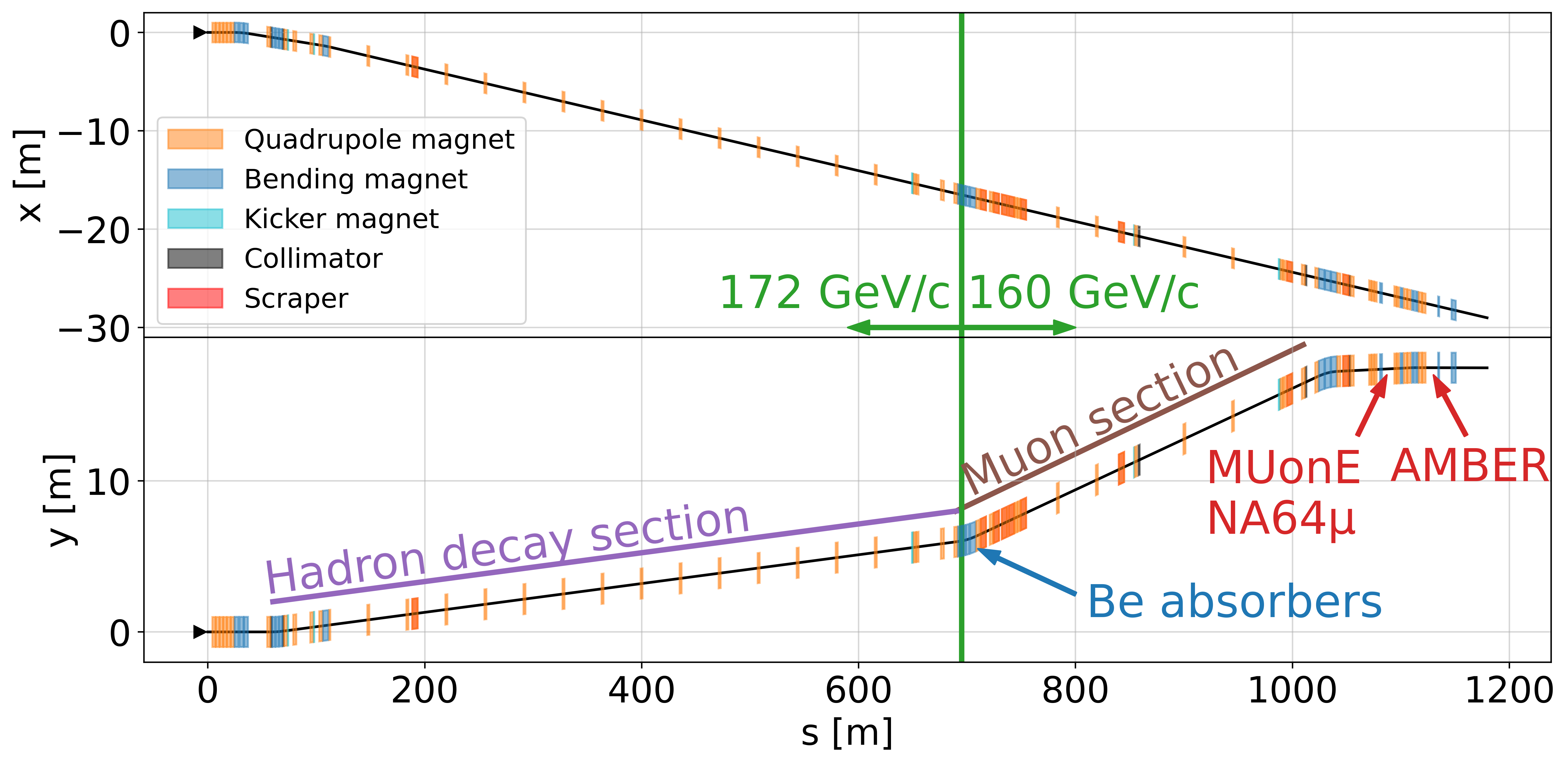}
    \caption{
        Schematic layout of the M2 beamline from the primary target (T6) towards the EHN2
        experimental hall. The hadron decay section (first vertical bend to second vertical bend) is indicated in purple.
        The muon section (second vertical bend to third vertical bend) is indicated in brown. A green
        line highlights the point where the central momentum definition changes to select the
        muon beam. The positions of the beryllium absorbers in the second vertical bend, of the MUonE and
        NA64$\upmu$ experiments, and of the AMBER experiment are also indicated.
    }
    \label{fig:m2layout}
\end{figure*}

The M2 beamline is a long, high-energy secondary beamline in the CERN North Area, served by
the T6 target station and designed to deliver hadron, electron, and muon beams to experiments
located in the EHN2 experimental hall. The beamline layout is shown in Fig.~\ref{fig:m2layout}.
The momentum selection, when operating the muon beam, is provided by three vertical bends, which recombine the vertical
dispersion and divide the beamline into four distinct sections: the target region up to approximately \SI{50}{\meter} from 
the target, a long (\SI{700}{\meter}) straight section in which muons are captured from hadron decays (hadron
decay section), a downstream straight section (\SI{400}{\meter}) long enough to control the
muon beam quality (muon section), and a final section where two experimental setups can be
installed.

\begin{figure*}
    \centering
    \includegraphics[width=\linewidth]{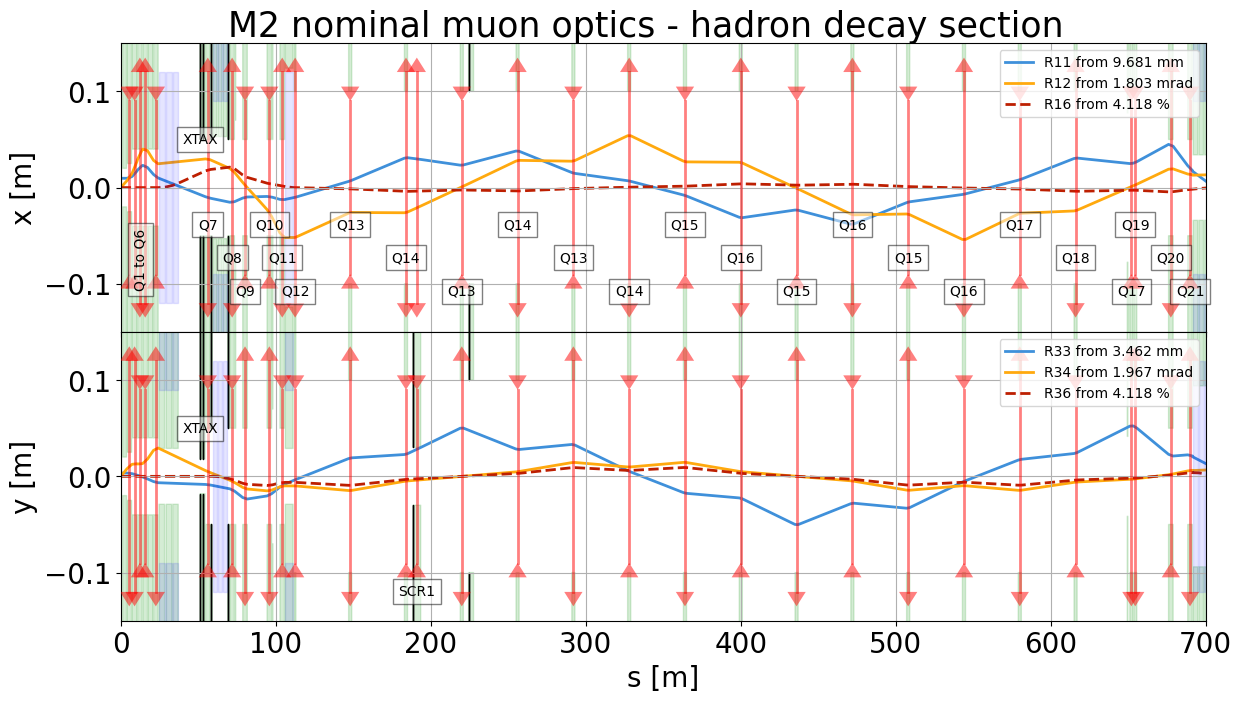}

    \vspace{0.5em}

    \includegraphics[width=\linewidth]{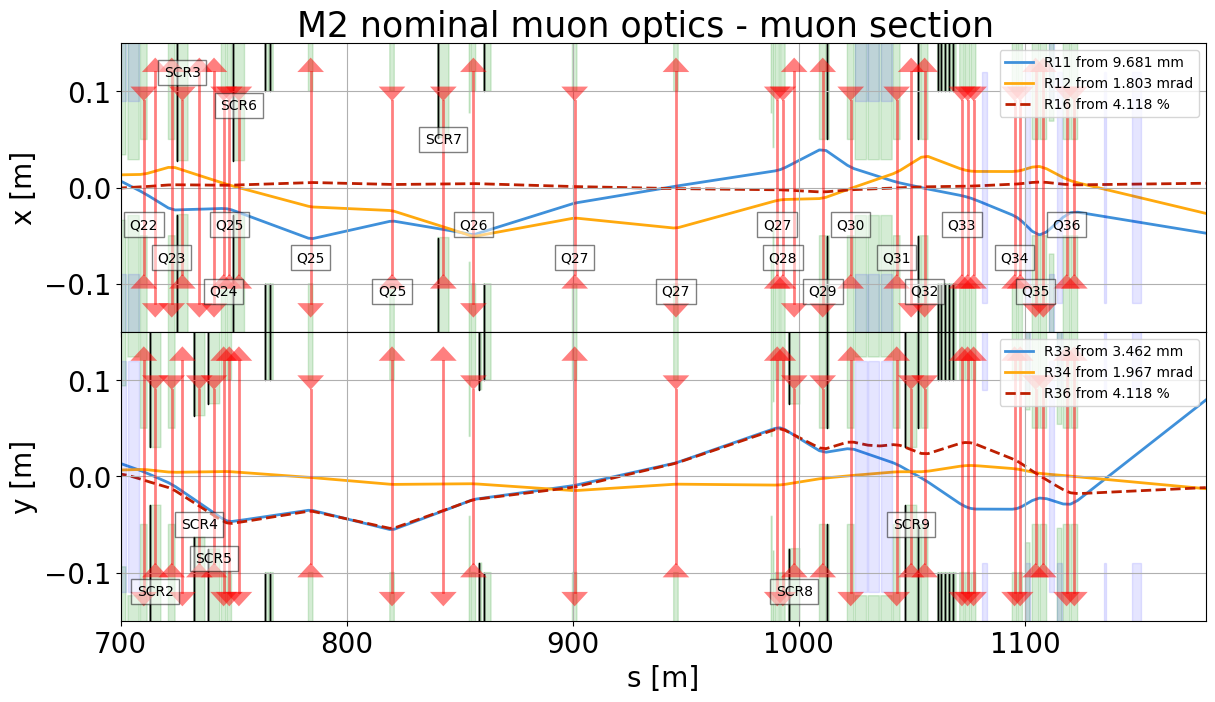}
    \caption{
        Nominal optics of the M2 beamline for muon operation along the hadron section (top) and the muon section (bottom).
        The R-matrix transport elements \cite{Brown:1968hkv} are shown for the horizontal (top) and vertical (bottom) planes, scaled
        to the maximum acceptance of the beamline.
        Quadrupoles (focusing/defocusing) are indicated by outward/inward double-headed red arrows,
        dipoles by blue boxes, apertures in green, and collimators as black lines.
        Quadrupoles with identical labels are powered in series by the same power converter.
        The labels of the scrapers are shown only in the plane they act on.
    }
    \label{fig:m2optics}
\end{figure*}

A first horizontal bend is used to increase the spatial distance between M2 and the neighboring P42 beamline,
and a second horizontal bend is used to recombine the horizontal dispersion.

The M2 lattice includes 48 quadrupole magnets up to the MUonE experiment, powered such that
the optics of the beamline is fully defined by 33 free parameters.

Depending on the operating configuration, M2 can deliver muon beams with momenta up to about
\SI{225}{\GeVoverc}, hadron beams up to \SI{280}{\GeVoverc}, and low quality electron
beams up to \SI{80}{\GeVoverc}.

\subsubsection{Muon beam production}
For muon operation, the beamline is configured such that the hadron decay section and the
downstream muon section operate at different central momenta. The decay channel is tuned to
transport hadrons at a higher momentum, while the downstream section is set to approximately
\SI{93}{\percent} of the hadron momentum. This choice represents a compromise between
maximizing the muon flux and achieving a high degree of muon polarization~\cite{LausNote}.
In this study, we focus on the nominal muon beam for the MUonE experiment, with a hadron
momentum of \SI{172}{\GeVoverc} and a muon momentum of \SI{160}{\GeVoverc}.

When delivering the muon beam, standard collimators cannot provide adequate control of the beam
phase space due to the limited interaction of high-energy muons with matter. Instead, a system of
nine toroidal magnetic collimators (scrapers), shown in red in Fig.~\ref{fig:m2layout}, is used to control the muon beam momentum spread and halo
by selectively deflecting particles outside the desired geometrical and momentum acceptance. The hadron
contamination in the muon beam is strongly reduced by nine \SI{1.1}{\meter} long beryllium rods, corresponding approximately
to \SI{28}{} radiation lengths and \SI{24.5}{} nuclear interaction lengths,
which can be mechanically inserted in the beam at the location of second vertical bend (see
Fig.~\ref{fig:m2layout}). When inserted, the hadron contamination is reduced to less than
\SI{e-5}{hadron/\upmu}~\cite{M2paper,Fabian}.

\subsubsection{Electron beam operation}
In contrast with H4, the XTAX collimator in M2 is located
downstream of six quadrupole magnets and the first horizontal bend, which provides the initial momentum
selection. Therefore, the production of secondary electron beams in M2 cannot rely on the
conversion of a collimated photon beam, since there is no direct line of sight between T6 and
the XTAX. Instead, electron beams are produced using a momentum degradation scheme in the
downstream part of the beamline.

In this mode of operation, the beam is first selected at the target at a momentum of about
\SI{100}{\GeVoverc}. A \SI{4}{\milli\meter} thick lead absorber is then inserted
\SI{50}{\meter} upstream of the second vertical bend (\SI{680}{\meter} from the T6 target), where electrons undergo significant energy loss, while hadrons and muons
lose comparatively little energy. The second vertical bend subsequently provides momentum selection of the
degraded electron component, allowing the extraction of electron beams with momenta up to about
\SI{80}{\GeVoverc}, though beam purity deteriorates at higher momenta.

Typical electron intensities obtained with this method are of the order of a few \SI{e4}{e^-}
per spill for \SI{3e12}{PoT} on the T6 target. 
In order to accommodate the scrapers, approximately \SI{100}{\meter} of the M2 beamline are not in vacuum,
which leads to significant degradation of the electron beam due to interactions with air.
As a result, the currently available electron beam is  unsuitable
for physics data-taking, but is sufficient for detector calibrations. Targeted upgrades for improving the beamline vacuum are expected to substantially improve
the achievable electron beam phase space and rate.

The rate, halo, and momentum acceptance of the M2 electron beam are controlled
by five collimators.

\subsubsection{Simulation input}
The input particle distributions for both the Xsuite and \textsc{BDSIM} models of the M2 beamline were
generated with a \textsc{BDSIM} simulation of the T6 target area. For the muon beam studies, a hadron
beam file was prepared by impinging \num{e7} protons on target. Both in Xsuite and \textsc{BDSIM}, the
hadron decay to muons was biased by a factor of 100 to reduce the variance of the Monte-Carlo sample.
Further details on the implementation of the hadron decay in the Xsuite model are provided in
the following section. This variance reduction technique is not expected to introduce significant statistical bias in the simulations due to the low efficiency of the muon production process and to the momentum selection. At \SI{172}{\GeVoverc} the decay length of pions, from which approximately \SI{90}{\percent} of the muon beam is generated, is \SI{9.93}{\kilo\meter}, so along the hadron section about \SI{7}{\percent} of the pions accepted by the first bend decay to muons. In addition, the resulting muons will be distributed uniformly in momentum between \SI{57}{\percent} and \SI{100}{\percent} of the pion momentum. For a \SI{160}{\GeVoverc} muon beam with a \SI{10}{\percent} momentum spread, the acceptance would be \SI{20}{\percent}, without accounting for the geometric acceptance of the beamline. With a muon production biasing factor of \num{100}, the final muon statistics is expected to be close to that of the hadron beam accepted by M2, and therefore the expected bias on the Monte Carlo sample is expected to be negligible.
This has been verified by running dedicated unbiased BDSIM simulations with \num{e9} protons on T6, finding agreement both in the delivered rates and the final beam phase spaces on the order of a few percent, consistent with the statistical variance of the produced sample.

Because the electron production yield is four orders of magnitude lower than the muon yield,
a dedicated simulation without muon biasing was run with \SI{e9}{} protons on target filtering
electrons only.

\begin{table*}
\caption{Benchmark at s=\SI{544}{\meter} of the Xsuite candidate solutions in H4 with \textsc{BDSIM}. Both models are evaluated in vacuum; the comparison including the full H4 material budget is shown in Fig.~\ref{fig:h4commissioning}. The uncertainties reported here represent only
the statistical significance of the samples.}
\label{tab:H4results}
\begin{ruledtabular}
\begin{tabular}{lccccc}
Optics & Rate [\SI{}{e^-/PoT}] & STD($x$) [\SI{}{\milli\meter}] & STD($y$) [\SI{}{\milli\meter}] & STD($x^\prime$) [\SI{}{\micro\radian}] & STD($y^\prime$) [\SI{}{\micro\radian}] \\
\hline
\multicolumn{6}{c}{Xsuite} \\
\hline
nominal & 2.65e-06 $\pm$ 1.63e-08 & 4.300 $\pm$ 0.017 & 11.544 $\pm$ 0.051 & 282.636 $\pm$ 1.056 & 108.538 $\pm$ 0.485 \\
optimized-like & 4.02e-06 $\pm$ 2.01e-08 & 4.365 $\pm$ 0.012 & 8.137 $\pm$ 0.025 & 376.737 $\pm$ 0.995 & 923.757 $\pm$ 2.670 \\
optimized-unlike & 4.36e-06 $\pm$ 2.09e-08 & 5.130 $\pm$ 0.014 & 4.847 $\pm$ 0.013 & 209.356 $\pm$ 0.637 & 372.068 $\pm$ 1.080 \\
\hline
\multicolumn{6}{c}{BDSIM - vacuum} \\
\hline
nominal & 2.65e-06 $\pm$ 1.63e-08 & 4.410 $\pm$ 0.029 & 11.741 $\pm$ 0.053 & 298.350 $\pm$ 1.407 & 110.593 $\pm$ 0.598 \\
optimized-like & 3.93e-06 $\pm$ 1.98e-08 & 4.327 $\pm$ 0.013 & 8.196 $\pm$ 0.026 & 384.244 $\pm$ 1.214 & 934.510 $\pm$ 2.766 \\
optimized-unlike & 4.39e-06 $\pm$ 2.10e-08 & 5.228 $\pm$ 0.015 & 4.983 $\pm$ 0.017 & 217.156 $\pm$ 0.718 & 403.145 $\pm$ 1.702 \\
\end{tabular}
\end{ruledtabular}
\end{table*}

\begin{figure*}
    \includegraphics[width=\linewidth]{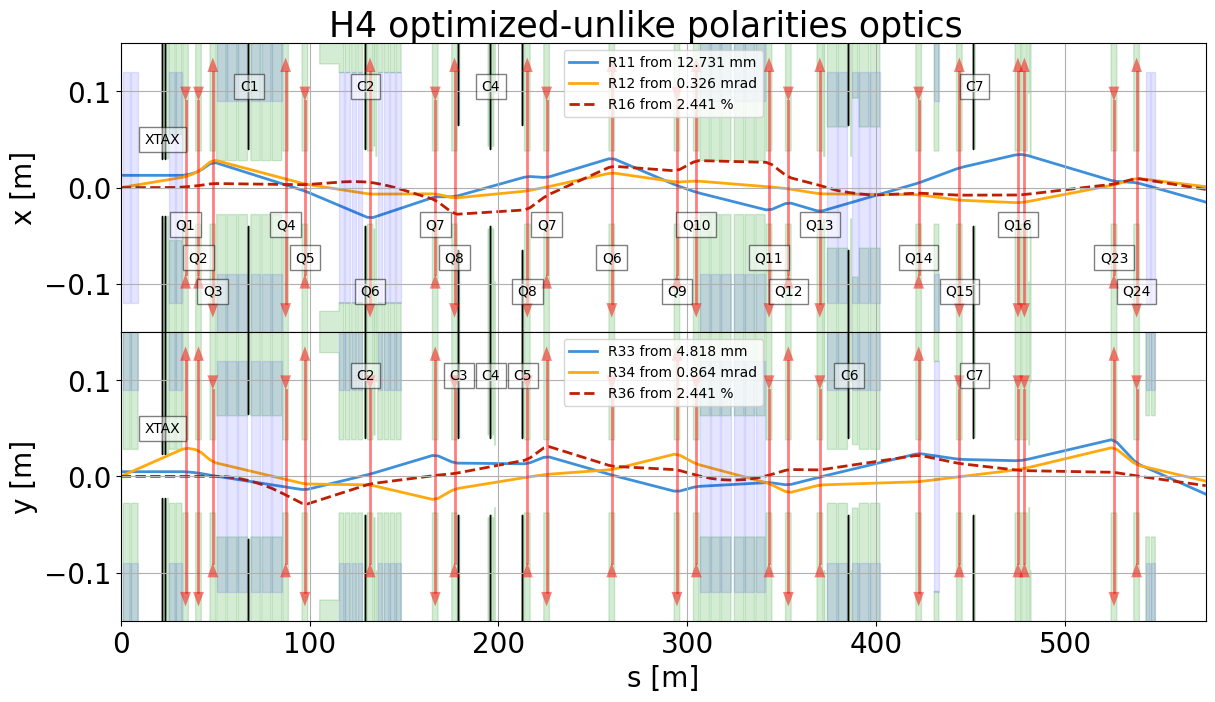}
    \caption{
        \textit{Unlike-polarity} optimized optics of the H4 beamline starting from the T2 target center.
        The R-matrix transport elements \cite{Brown:1968hkv} are shown for the horizontal (top) and vertical (bottom) planes, scaled
        to the maximum acceptance of the beamline.
        Quadrupoles (focusing/defocusing) are indicated by outward/inward double-headed red arrows, dipoles by blue boxes, apertures in green, and collimators as black lines.
        Quadrupoles with identical labels are powered in series by the same power converter. The labels of the collimators are shown only in the plane they act on.
    }
    \label{fig:h4optimizedoptics}
\end{figure*}

\section{Optimization}
The optimization workflow is based on fast tracking models implemented in Xsuite.
For both H4 and M2, the optics models were benchmarked against the corresponding existing
optics simulations used in operation.

The optimization is performed in two separate and subsequent stages. In the first one, the
\textit{mono-objective stage}, all quadrupoles are optimized to maximize the transmitted rate
at the experimental location. The aim is to find the settings that yield the highest rate, independently
of the final beam phase space. 
At the end of this stage, the settings of the quadrupoles up to
a breaking point far upstream in the beamline are fixed, thus reducing the number of free parameters for the
following stage: Q13 chosen in H4 (see Fig.~\ref{fig:h4layout}) and Q21 in M2 (see Fig.~\ref{fig:m2layout})\footnote{In M2, Q21 represents a natural breaking point, as it separates the upstream hadron section and the downstream muon section. In H4, such a clear separation does not exist and Q13 was chosen in order to have one more degree of freedom than the number of figures of merit.}.
A new particle distribution is generated with a high-fidelity \textsc{BDSIM} simulation up to that same point.

The \textit{multi-objective stage} then optimizes the downstream section starting from this distribution: the
rate is maximized while simultaneously minimizing the transverse beam size and divergence at the experimental
location, with all figures of merit and constraints evaluated at the experiment.
Beam size and divergence are quantified as the standard deviations of the marginal distributions in
$(x,y)$ and $(x',y')$.

The experiment-specific requirements are encoded as constraints on the figures of merit, and are specified for each
beamline in the following sections. Figures of merit are the quantities being optimized --- rate, beam size,
divergence --- while constraints are selection rules that restrict the exploration to feasible and physically
acceptable regions of that space. Together, they provide a compact and modular way to steer the same optimizer
towards different experimental goals.

We make use of the NSGA-II algorithm~\cite{NSGA-II,NSGAII}, a population-based evolutionary algorithm well suited to
high-dimensional, non-convex, and multi-objective problems. To accelerate early convergence, the population
is initialized using a Tree-structured Parzen Estimator (TPE) Bayesian optimizer~\cite{TPE}, as implemented in the
\texttt{NSGAIIWithTPEWarmupSampler} available in the OptunaHub~\cite{NSGAIIWithTPEWarmupSampler} repository. The TPE samples promising
regions of the parameter space based on the history of previous evaluations, providing the evolutionary algorithm with
a better-than-random starting population. Because random initial samples frequently yield non-transmitting optics,
this \textit{warm-up} stage is continued until the initial population contains only trials transmitting at least one particle
downstream of the breaking point. Both the NSGA-II algorithm and the TPE sampler are accessed through the Optuna hyperparameter
optimization framework~\cite{Optuna}.

\begin{figure*}
    \includegraphics[width=\linewidth]{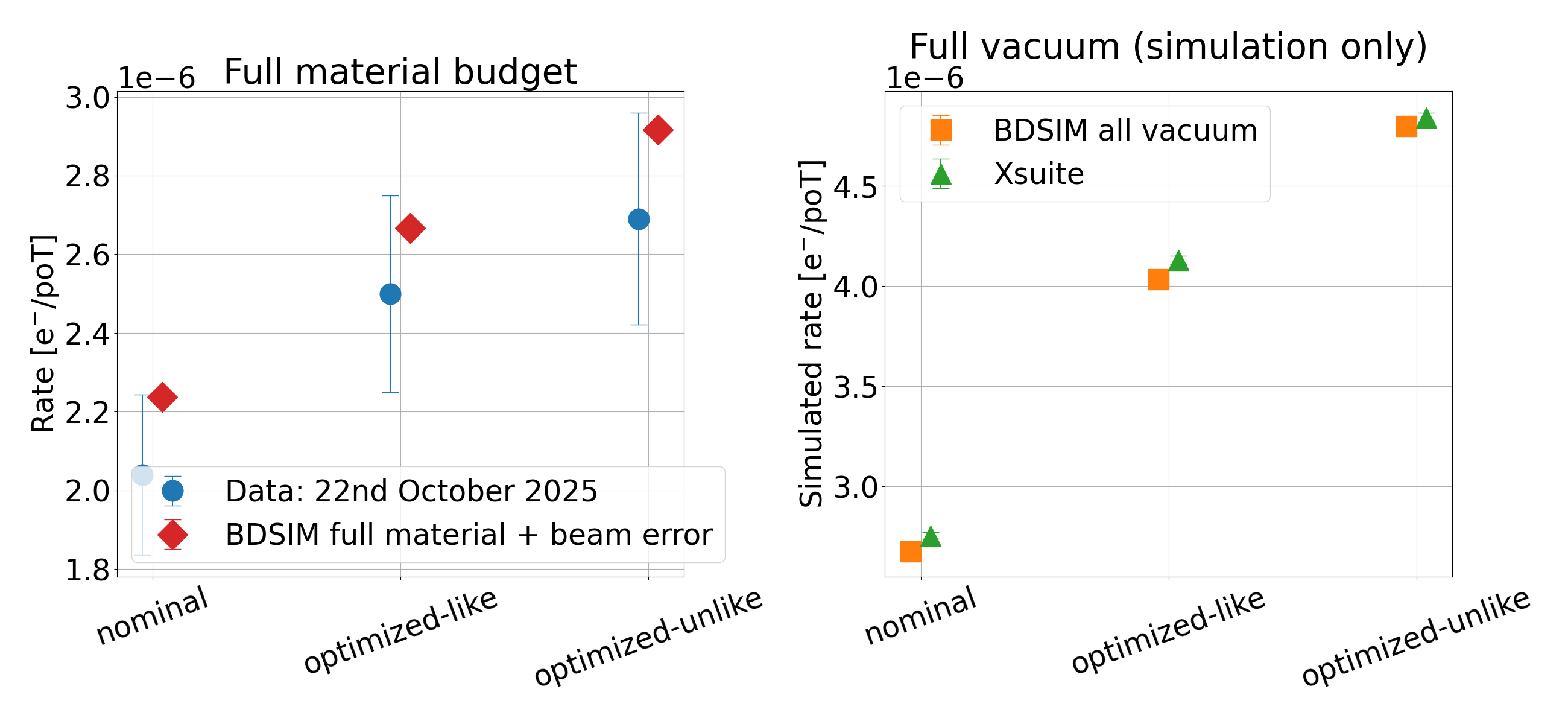}%
    \caption{
        Electron rates at the last scintillating detector before the NA64 experiment ($s=\SI{481}{\meter}$ in H4), normalized to the number of protons on target (PoT), for the nominal and the two optimized optics. \textit{Right}: simulation-only comparison in vacuum between \textsc{BDSIM} (orange) and Xsuite (green), which agree at the few-percent level. \textit{Left}: measured rates (blue, recorded on 22 October 2025) compared with the \textsc{BDSIM} model including the full H4 material budget and a \SI{200}{\micro\radian} error on the initial electron beam direction (red), representative of the commissioning conditions. This configuration reproduces the absolute measured rates within \SI{10}{\percent} and the relative increase between optics within the measurement uncertainty. Only the statistical uncertainty on the simulated rates is shown; the error bars on the measured rates represent an estimated \SI{10}{\percent} systematic uncertainty (see text).
    }
    \label{fig:h4commissioning}
\end{figure*}

\begin{figure*}
    \includegraphics[width=\linewidth]{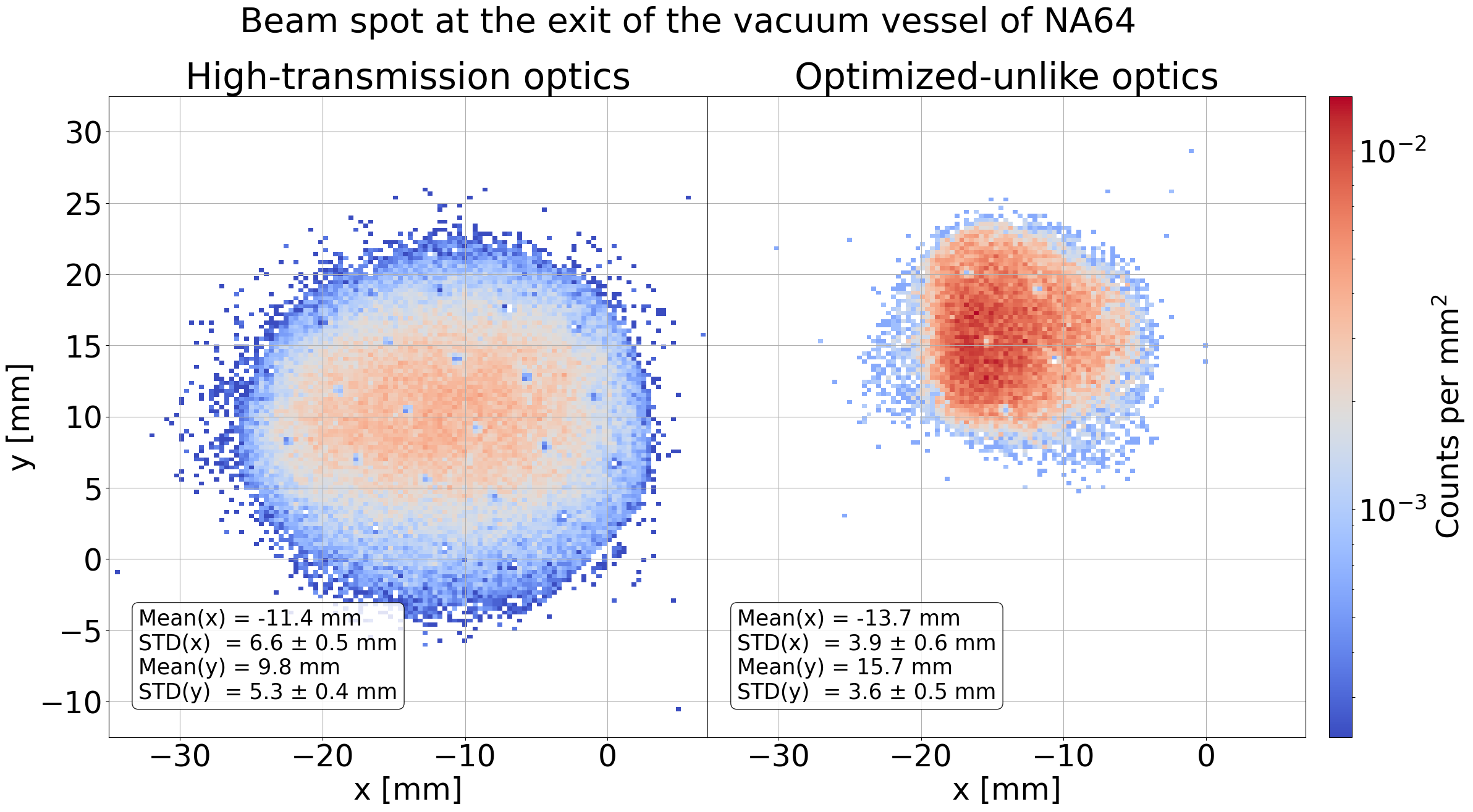}
    \caption{
        Beam spot at the exit of the NA64 vacuum vessel ($s=\SI{559}{\meter}$, see Micromegas 3 detector in~\cite{NA642023168776}) measured with the
        nominal optics (left) and with the optimized \textit{unlike-polarity} optics (right). The reduction in
        beam size is achieved while increasing the electron rate delivered to the experiment by \SI{30}{\percent}
        and while the beam halo content is reduced from \SI{5}{\percent} to \SI{1}{\percent}. Courtesy of
        the NA64 collaboration.
    }
    \label{fig:h4na64beamspots}
\end{figure*}

\subsection{H4 optimization}
To explore the full parameter space of the \textit{mono-objective stage}, the quadrupoles should in principle be allowed to
assume either polarity (N or S), as allowed by their power supplies without any restriction. However, from dedicated test runs, including both polarities simultaneously in a single optimization run increases the size of the search space to the point that the algorithm fails to converge to solutions with non-zero beam transmission\footnote{This is due to a combination of the so-called \textit{curse of dimensionality},
so to say the exponential increase of the parameter space with the dimensionality of the problem, and to the fact that beam transmission is a sparse function. Intuitively, for any additional parameter with allowed polarity
change, the number of local maxima in transmission doubles, increasing the sparsity of the problem with its dimensionality.}. To mitigate
this, the \textit{mono-objective stage} was performed twice under different assumptions on the polarity of the quadrupoles in the
upstream section (Q1 to Q12, see Fig.~\ref{fig:h4layout}). Q13, the last quadrupole before the second vertical bend, is included in the
\textit{multi-objective stage} together with the five downstream quadrupoles of the final focusing section (see Fig.~\ref{fig:h4optimizedoptics}).
The first case, referred to as \textit{like-polarity} optimization, constrains the upstream quadrupoles to retain the same polarity
as in the nominal high-transmission optics. The second case, referred to as \textit{unlike-polarity} optimization, constrains them
instead to assume the opposite polarity.

The \textit{multi-objective stage} was guided by defining a set of constraints aimed at strictly improving the beam spot size and divergence at the experiment with respect to the nominal optics:
\begin{itemize}
    \item $\sqrt{\mathrm{STD}(x)^2 + \mathrm{STD}(y)^2} < \SI{20}{\milli\meter}$.
    \item $\mathrm{STD}(x) < \SI{5.6}{\milli\meter}$ ($1.3 \times \mathrm{STD}(x)_\text{nominal}$).
    \item $\mathrm{STD}(y) < \SI{15}{\milli\meter}$ ($1.3 \times \mathrm{STD}(y)_\text{nominal}$).
    \item $\mathrm{STD}(x) < \mathrm{STD}(y)$, as delivered by the nominal optics.
    \item Acceptance $> \SI{7.8}{\percent}$: fraction of transmitted electrons within the range $[\SI{95}{\GeVoverc}, \SI{105}{\GeVoverc}]$, to be compared with the nominal optics acceptance of \SI{6.8}{\percent}. The threshold is chosen arbitrarily to require at least a \SI{15}{\percent} improvement, but, as for the other constraints, it should be considered as a mean to bias the exploration, rather than an exclusion principle for possible solutions. During the optimizations, the full history is stored for later analysis, including the evaluations that do not satisfy the constraints. As an example, the point chosen for the \textit{unlike-polarity} violates the constraint requiring the vertical beam size to exceed the horizontal one (see Table~\ref{tab:H4results}).
\end{itemize}
The first four constraints serve to focus the exploration on solutions with a beam spot size comparable to or smaller than that delivered to the NA64 detector with the nominal optics. The last constraint ensures that higher rate solutions are favored.

All figures of merit (rate, beam spot size and divergence) and constraints are evaluated at a distance $s=\SI{544}{\meter}$ from T2, at the center of the spectrometer dipole of the NA64 experiment (see Fig.~\ref{fig:h4layout}).

At the end of the \textit{mono-objective stage}, the \textit{like-polarity} solution yielded a maximum acceptance of \SI{11.6}{\percent}, and the \textit{unlike-polarity} solution yielded \SI{15.7}{\percent}, respectively 1.7 and 2.3 times higher than the baseline nominal optics acceptance of \SI{6.8}{\percent}. Two candidate configurations were then selected from the best trade-off points, or Pareto fronts, of the multi-objective optimization, defined as the highest rate solutions with a beam spot size smaller than
\SI{6}{\milli\meter} and divergence smaller than \SI{1}{\milli\radian} in both planes. The solutions were
benchmarked with the high-fidelity \textsc{BDSIM} model. The results are summarized in Table~\ref{tab:H4results}.
The uncertainties reported represent only the statistical significance of the samples. The two models show
agreement at the level of a few percent.
The \textit{unlike-polarity} solution achieves the highest rate, with the smallest beam spot size and divergence,
and was therefore chosen for the tests with the NA64 detector. Figure~\ref{fig:h4optimizedoptics} shows the \textit{unlike-polarity}
optics diagram. In comparison to the nominal optics (Fig.~\ref{fig:h4optics}), the geometric acceptance is reduced from
\SI{0.37}{\pi\micro\steradian} to \SI{0.28}{\pi\micro\steradian}, while the momentum acceptance is increased from
\SI{1.4}{\percent} to \SI{2.4}{\percent}, resulting in a net increase of the accepted phase space. The limiting aperture in the horizontal plane moves from Q5 to the first vertical bend, and in the vertical plane it moves from the kicker magnet located around $\mathrm{s=\SI{200}{\meter}}$ to the XTAX.

As shown in Table~\ref{tab:H4results}, the nominal optics delivers a large vertical beam spot
($\mathrm{STD}(y) \approx \SI{11}{\milli\meter}$).
In this configuration, the vertical jaws of collimator C4 (located at $s \approx \SI{200}{\meter}$, see
Fig.~\ref{fig:h4layout}) must be partially closed to reduce the beam size prior to delivery to the experiment.
In addition, the vertical jaws of C2, positioned at a point of maximum dispersion (see
Fig.~\ref{fig:h4optics}), provide the primary control of the momentum acceptance.

In the \textit{unlike-polarity} optics, the final beam spot at the experimental location is already within the range required by NA64,
so C4 is no longer used to compensate for beam size. In general, due to the different location of the focal points in the new optics,
the overall collimation scheme changes (see Table~\ref{tab:H4collimation}).
Rate control, which in the nominal configuration relied on C1 together with the horizontal jaws of C4,
can now be achieved primarily with C2. Conversely, the role of momentum selection shifts from the vertical jaws of C2
to the horizontal jaws of C4. Due to the lack of a focal point with a maximum dispersive component, this configuration provides somewhat less effective momentum selection.
On the other hand, it does not require, in principle, collimation to reduce the beam size, thus allowing for higher rates to be delivered to
the experiment.

\subsubsection{Results}
The two new optics configurations were tested during commissioning in April 2025 and dedicated tests in October 2025, to be compared with the nominal optics for validation of the models.
The test was performed with the \SI{500}{\milli\meter} long beryllium target and with all the collimators open to
\SI{\pm40}{\milli\meter}.
Figure~\ref{fig:h4commissioning} shows the measured electron rates normalized to the number of protons on target
(PoT) per spill at the last scintillating detector in H4 before the NA64 experiment, placed at $s=\SI{481}{\meter}$.

\begin{table}
\caption{H4 collimation scheme for the nominal high-transmission optics and the \textit{unlike polarities} optimized optics.}
\label{tab:H4collimation}
\begin{ruledtabular}
\begin{tabular}{lcc}
 & Rate control   & Momentum selection\\
\hline
nominal           & C1 and C4 horizontal & C2 vertical\\
unlike-polarities & C2                   & C4 horizontal\\
\end{tabular}
\end{ruledtabular}
\end{table}

In vacuum, the \textsc{BDSIM} and Xsuite models agree at the few-percent level (Fig.~\ref{fig:h4commissioning}, right), but both overestimate the measured rates. This is expected, since during the commissioning run the beamline vacuum was not yet fully established. Reproducing those conditions in \textsc{BDSIM}, by including the full H4 material budget, lowers the simulated rate by \SI{20}{\percent}; the remaining difference is recovered with a \SI{200}{\micro\radian} error on the initial direction of the electron beam at T2 (Fig.~\ref{fig:h4commissioning}, left). With both effects included, the simulation reproduces the absolute rates within \SI{10}{\percent} and the relative increase between the nominal and the optimized optics within the measurement uncertainty.

The beam-steering error is not the result of a fit and does not represent the only source of discrepancy. It is introduced to show that a misalignment comparable to the accuracy of the beam steering onto T2 is sufficient to account for the observed rates. Further systematic contributions are expected from the positioning of the beamline elements and monitors, which cannot be measured directly in the target area, and from the limited knowledge of the detector response and efficiency, the latter accounting for the \SI{10}{\percent} uncertainty assigned to the measured rates. Given these effects, the predicted absolute rates and relative increase are consistent with the experimental observation.

%The two simulation models (\textsc{BDSIM} and Xsuite) show good agreement and overestimate the absolute measured rates
%by a factor $\sim2$.
%The discrepancy can be attributed to effects not included in the simulations, such as misalignments of the beamline elements, which are known to be worse than \SI{1}{\milli\meter} depending on the location,
%and inaccuracies in the modeling of the beamline apertures. The origin of this discrepancy is currently under investigation.
%However, the relative increase in rate measured with the new optics settings is compatible with the expectation. For the
%scope of this study, this is sufficient in order to validate the optimization approach and the models used.

An additional test was performed with the NA64 collaboration to compare the performances of the nominal
optics and of the \textit{unlike-polarity} optics. Under these conditions,
the veto counter at the entrance of the spectrometer magnet (see \cite{NA642023168776} for reference) recorded a five-fold reduction of the
beam halo from \SI{5}{\percent} to \SI{1}{\percent} due to the reduction of the beam spot size.
Fig.~\ref{fig:h4na64beamspots} shows a comparison of the beam spots at the exit of the NA64 vacuum
vessel, measured with the nominal and \textit{unlike-polarity} optics.

\subsection{M2 optimization}
The optimization strategy of the M2 beamline follows closely that of H4. The \textit{mono-objective stage} maximizes
the transmitted muon rate at the MUonE experiment location, at $s=\SI{1079}{\meter}$, as a function of all the quadrupole strengths.
The Xsuite M2 model is used here to propagate hadrons from T6 through the hadron decay section
and includes their decays into muons, as well as the interaction of muons with the beryllium absorbers in the second vertical bend. Further
details are provided in the dedicated paragraph below. The polarities of the quadrupoles in the hadron decay section
(up to Q21, see Fig.~\ref{fig:m2layout}) are fixed to match the nominal optics, while the quadrupoles in the muon section can assume either polarity.

To avoid biasing the exploration of the parameter space, in the \textit{mono-objective stage} all scrapers
except the first one --- which defines the overall acceptance of the beamline --- are kept fully open, and their
apertures are introduced as parameters in the \textit{multi-objective stage}, when optimizing the
muon section.

Due to the larger parameter space of 33 quadrupole strengths (to be compared with 18 in H4), the \textit{mono-objective stage} already includes constraints, evaluated at the experiment location, to drive the exploration towards solutions with acceptable transverse phase-space properties at the MUonE experiment:
\begin{itemize}
    \item $\sqrt{\mathrm{STD}(x)^2 + \mathrm{STD}(y)^2} < \SI{30}{\milli\meter}$ ($\lessapprox$ nominal).
    \item $\mathrm{STD}(x^\prime) < \SI{1}{\milli\radian}$ ($2 \times \mathrm{STD}(y^\prime)_\text{nominal}$).
    \item $\mathrm{STD}(y^\prime) < \SI{1}{\milli\radian}$ ($2 \times \mathrm{STD}(y^\prime)_\text{nominal}$).
\end{itemize}
At the end of the \textit{mono-objective stage}, the settings of the hadron decay section are fixed to the best solution
found. The muon rate at the entrance of the second vertical bend, before the beryllium absorbers, is increased by \SI{35}{\percent} with
respect to the nominal optics, reflecting the improved overall acceptance of the hadron decay section. A dedicated \textsc{BDSIM}
simulation is then used to generate a muon beam at the exit of the second vertical bend, after the absorbers, at $s=\SI{708}{\meter}$
from T6.

The \textit{multi-objective stage} optimizes the strengths of the quadrupoles and the scraper apertures in the muon section,
from the second vertical bend to the MUonE experiment, maximizing the muon rate while minimizing the beam size at the experiment. Since the
input distribution is the muon beam at the exit of the second vertical bend as generated by \textsc{BDSIM}, only the optics of the muon section
is required in this stage. The constraints applied here are more stringent than in the \textit{mono-objective stage}{, to favor the exploration closer to a strict improvement with respect to the nominal}:
\begin{itemize}
    \item $\sqrt{\mathrm{STD}(x)^2 + \mathrm{STD}(y)^2} < \SI{20}{\milli\meter}$ ($\leq \SI{70}{\percent}$ nominal).
    \item $\mathrm{STD}(x^\prime) < \SI{0.7}{\milli\radian}$ ($1.3 \times \mathrm{STD}(y^\prime)_\text{nominal}$).
    \item $\mathrm{STD}(y^\prime) < \SI{0.7}{\milli\radian}$ ($1.3 \times \mathrm{STD}(y^\prime)_\text{nominal}$).
    \item $g_1(x) < 0.2$, where $g_1$ is the Fisher-Pearson sample skewness~\cite{statistics}, to favor symmetric beam profiles.
    \item $g_1(y) < 0.2$.
    \item $g_1(p) < 0.1$.
\end{itemize}
The constraints on skewness were not necessary for H4, as its momentum acceptance ($\sim \SI{2}{\percent}$) is much narrower
than for M2 ($\sim \SI{10}{\percent}$ for hadrons). During preliminary studies, it was observed that a fraction of otherwise feasible
solutions delivered a highly asymmetric beam at the experiment, with a long tail in momentum and correspondingly asymmetric
beam spots due to higher order dispersive effects.
As mentioned in the definition of the constraints for the optimization of H4, the constraints are used to bias the exploration towards a class of solutions, rather than an exclusion principle.

The final solution selected for the M2 muon optics delivers a tighter focus at the experiment, reducing the vertical beam size standard deviation from \SI{35}{\milli\meter} to \SI{18}{\milli\meter},
together with a factor-of-two increase in rate with respect to the nominal optics. This substantial improvement in rate is primarily due to the inclusion
of the scrapers in the optimization, making it possible to accept larger apertures while still meeting the experimental requirements on
the beam phase space.
Table~\ref{tab:M2results} shows a comparison between the nominal and optimized beam parameters at the MUonE experiment location,
as simulated with Xsuite.

\begin{table*}
\caption{Comparison between the beam parameters of the nominal and optimized M2 muon optics at the MUonE experiment location,
$s=\SI{1079}{\meter}$ from T6, as simulated with Xsuite. The uncertainties reported here represent only the statistical
significance of the simulated Monte-Carlo samples, weighted for different reduction techniques applied.}
\label{tab:M2results}
\begin{ruledtabular}
\begin{tabular}{lcccccc}
Optics & Rate [\SI{}{\upmu^+/PoT}] & STD($x$) [\SI{}{\milli\meter}] & STD($y$) [\SI{}{\milli\meter}] & STD($x^\prime$) [\SI{}{\micro\radian}] & STD($y^\prime$) [\SI{}{\micro\radian}] & STD(p) [\SI{}{\GeVoverc}] \\
\hline
nominal          & (1.377 $\pm$ 0.011)$\times \SI{e-5}{}$ & 10.57 $\pm$ 0.05 & 34.62 $\pm$ 0.18 & 331.8 $\pm$ 1.8 & 568 $\pm$ 3 & 5.37 $\pm$ 0.03\\
optimized        & (2.714 $\pm$ 0.016)$\times \SI{e-5}{}$ & 10.40 $\pm$ 0.04 & 17.82 $\pm$ 0.08 & 502.3 $\pm$ 1.8 & 777 $\pm$ 4 & 6.12 $\pm$ 0.02\\
\end{tabular}
\end{ruledtabular}
\end{table*}

\subsubsection{Fast muon phase-space modeling in M2}
The muons in M2 are captured along the \SI{\sim700}{\meter}-long hadron decay section (see Fig.~\ref{fig:m2layout}), carrying between
\SI{57}{\percent} and \SI{100}{\percent} of the parent hadron momentum, making the parametrization of an initial muon distribution a non-trivial task.
A more natural approach consists in modeling the production of the muons from the decays of hadrons
propagated from the T6 target. This required a minimal extension of the optics-only Xsuite model
through the introduction of the \texttt{BeamInteraction} elements, readily available in Xsuite and originally implemented
to model the interaction with collimator jaws \cite{Xsuite}.
The hadron decay was modeled as an element with no length, placed at regular intervals of \SI{20}{\meter} along the
decay channel. The choice of spacing was verified to have negligible impact on the resulting muon distribution.
At each of such elements, each hadron is checked as possible candidate to decay based on its decay length and on
the space travelled from the previous \texttt{BeamInteraction} element. In order to reduce the variance of the Monte-Carlo sample,
the hadron decay can be biased to produce multiple muons per parent hadron, adjusting their weights accordingly \cite{10.3389/fphy.2021.718873}. For the optimization studies presented
here, the bias factor was set to 10, meaning that each hadron decay produces up to 10 muons along the decay channel.
To speed up the tracking, the hadron beam is first propagated up to the beryllium absorbers without decays enabled.
The surviving hadrons are then retracked from the target with decays enabled, producing the muon beam, thus reducing the number
of tracks traversing the \texttt{BeamInteraction} elements.
An analogous biasing of the muon production is applied in the \textsc{BDSIM} simulations used for the benchmark comparisons.

Energy loss and scattering in the beryllium absorbers located in the second vertical bend are modeled analogously. Each of the three
\SI{3}{\meter} long dipole magnets composing the bend is sliced into ten shorter magnets, interleaved with
\texttt{BeamInteraction} elements (11 per magnet) that apply energy loss, straggling, and multiple Coulomb scattering
from \SI{3.3}{\meter} of beryllium per dipole.
Because high-energy muons are highly penetrating, they are not removed by
standard non-magnetic apertures; therefore, when tracking the surviving hadron beam with the
additional physics enabled, non-magnetic apertures, i.e. beam pipes, non-magnetic collimators
and beam instrumentation, are removed from the model.

\subsubsection{Results}

The optimized M2 muon optics were tested during a dedicated commissioning run in 2025, performed with the
\SI{500}{\milli\meter} long beryllium target in T6 and with all non-magnetic collimators fully open. The muon
rate was measured with an ionization chamber located at $s=\SI{1093}{\meter}$ from T6.
The muon rate per proton on target was measured for the nominal and optimized optics, each tested with both the nominal
and the optimized scraper settings in order to disentangle their contribution to the rate improvement.
Table~\ref{tab:scrapers} shows the nominal and optimized half-apertures of the scrapers in the muon section. The optimization
yielded different apertures for SCR3, SCR6, SCR7, and SCR9, while the other scrapers were left unchanged.
The results are summarized in Table~\ref{tab:M2commissioning}.

\begin{table}[h]
\caption{Nominal and optimized scraper half-apertures in M2 (see Fig.~\ref{fig:m2optics}). Scrapers whose aperture
was not changed by the optimization are shown for completeness.}
\label{tab:scrapers}
\begin{ruledtabular}
\begin{tabular}{lccc}
Scraper & Plane & Nominal [\SI{}{\milli\meter}] & Optimized [\SI{}{\milli\meter}] \\
\hline
SCR1 & Vertical   & 30.0  & 30.0  \\
SCR2 & Vertical   & 30.0  & 30.0  \\
SCR3 & Horizontal & 28.0  & 65.0  \\
SCR4 & Vertical   & 63.0  & 63.0  \\
SCR5 & Vertical   & 75.4  & 75.4  \\
SCR6 & Horizontal & 28.0  & 40.0  \\
SCR7 & Horizontal & 51.9  & 74.0  \\
SCR8 & Vertical   & 75.1  & 75.1  \\
SCR9 & Vertical   & 30.0  & 50.0  \\
\end{tabular}
\end{ruledtabular}
\end{table}

\begin{table}[h]
\caption{Measured muon rates at $s=\SI{1093}{\meter}$ from T6 for the four commissioning configurations tested during
the 2025 run.}
\label{tab:M2commissioning}
\begin{ruledtabular}
\begin{tabular}{lcc}
    & \multicolumn{2}{c}{\textbf{Rates in \SI[detect-weight=true]{e-5}{\upmu^+/\text{PoT}}}} \\
    \hline
    & Nominal scrapers & Optimized scrapers \\
    \hline
    Nominal optics   & 1.47 &  1.89  \\
    Optimized optics & 1.96 &  2.46  \\
\end{tabular}
\end{ruledtabular}
\end{table}

The fully optimized configuration --- optimized optics combined with optimized scraper apertures --- yielded a muon rate of
\SI{2.46e-5}{\upmu^+/PoT}, corresponding to a \SI{67}{\percent} increase with respect to the nominal optics with nominal scraper
settings, which delivered \SI{1.47e-5}{\upmu^+/PoT}. The intermediate configurations show that the optics and the apertures of
the scrapers separately contribute a \SI{\sim 30}{\percent} increase each. It should be noted that the optimized scraper apertures are specific to the MUonE experiment, which imposes no stringent constraints on beam halo. Experiments such as COMPASS and AMBER, which share the M2 beamline, require tighter halo control, and the nominal scraper settings reflect this more stringent operating condition.
In contrast to H4, the simulated relative increase in muon rate reproduces the measured one without introducing any beam-steering error. This is expected, given the much larger acceptance of M2: a mis-steering of \SI{2.5}{\milli\radian} on the target reduces the muon rate by only \SI{10}{\percent}. A mis-steering of the magnitude relevant for the electron beam therefore has a negligible effect on the muon rate, consistent with the good agreement between the simulated and measured rates.

In addition, an optimization run was performed on the first six quadrupoles of the M2 beamline to
increase the transmission of the electrons produced at T6, through the XTAX. Due to the low dimensionality of the problem,
the quadrupoles were allowed to assume either polarity. The new settings were tested during a
dedicated commissioning run in 2025, measuring \SI{1.23e-08}{e^-/PoT} at the scintillating detector
located at $s=\SI{1079}{\meter}$ from T6. Compared with the nominal optics, which delivered
\SI{6.74e-09}{e^-/PoT}, this optimization resulted in a \SI{82}{\percent} increase in electron rate.
The new electron optics were deployed to the MUonE experiment for the calibration run of their electromagnetic
calorimeter.

\subsection{Computing performance}
All the optimizations presented in this work were run on a standard workstation in parallel on 8 cores.
This was made possible by the implementation of the Xsuite based models, which allowed to 
perform a large number of evaluations with limited computing resources.

The required computing time for a high-fidelity \textsc{BDSIM} simulation depends on the level of detail
and on the figures of merit to be evaluated. In order to compare with the Xsuite based models,
a \SI{10}{\GeVoverc} minimum kinetic energy cut was applied in \textsc{BDSIM}, to prevent
the tracking of low-energy secondary particles that would not contribute to the propagated beam.
For the comparison of the M2 muon beam, the two models have been run with the same muon biasing factor
of 100.

Table~\ref{tab:computingperformance} summarizes the wall time per hadron and per electron tracked in M2
and H4, respectively, for both the Xsuite and \textsc{BDSIM} models.

The resulting speed-up relative to high-fidelity \textsc{BDSIM} tracking is \SI{e3}{} for
H4 and \SI{e2}{} for M2, despite the inclusion of hadron decays and material interactions in the absorbers.

\begin{table}[h]
\caption{Wall time per hadron/electron tracked in M2 and H4 for both the Xsuite and \textsc{BDSIM} models.}
\label{tab:computingperformance}
\begin{ruledtabular}
\begin{tabular}{lcc}
Beamline & Xsuite [ms/particle] & \textsc{BDSIM} [ms/particle] \\
\hline
H4 (e$^-$) & 0.003 & 3 \\
M2 ($\uppi^+/\mathrm{K}^+$)   & 0.15  & 18 \\
\end{tabular}
\end{ruledtabular}
\end{table}
\section{Conclusions}

We have presented the optimization of the H4 and M2 beamlines in the CERN North Area using
large-scale optimization algorithms coupled to fast low-fidelity tracking models.
High-fidelity simulations were reserved for benchmarking the final candidate solutions.
The impact of the material budget and of errors in the element alignment and
the initial beam position was also assessed, showing good agreement with the experimental observations.
The use of fast models reduces the computational cost to the point that the full optimization can
be run on a standard workstation, enabling the exploration of high-dimensional parameter
spaces that would otherwise be impractical. The approach was demonstrated on the electron
beam of H4 and on both the muon and electron beams of M2, covering a range of beam
compositions and physical processes of increasing complexity. This approach is
completely general and its application is not limited to secondary beamlines, with the main
requirement being the possibility to perform tens of thousands of evaluations for each optimization run.
For H4, new electron optics were deployed to the NA64 experiment during the 2025 run,
resulting in a \SI{30}{\percent} increase in electron rate and a five-fold reduction in
beam-halo content, from \SI{5}{\percent} to \SI{1}{\percent}. This was achieved by relaxing
the strict momentum selection of the nominal high-transmission optics in favor of increased
momentum acceptance, while simultaneously tightening the focus at the entrance of the NA64
spectrometer.

For M2, the optimization of the muon beam required extending the fast tracking model to
include hadron decays in the decay section and muon energy loss and scattering in the
beryllium absorbers at the second vertical bend. Despite the increased complexity, the optimized muon optics
was successfully commissioned in 2025, yielding a \SI{67}{\percent} increase in muon
transmission. A separate optimization of the electron transmission through the M2 XTAX was performed on the first six quadrupoles of M2,
yielding an \SI{82}{\percent} increase in electron rate, and the resulting optics
were deployed for the MUonE electromagnetic calorimeter calibration run in 2025.

Looking ahead, vacuum consolidation works are expected to substantially reduce beam losses
along M2, opening the prospect of significantly higher electron beam rates in future runs.
For the muon beam, radiation protection studies are currently ongoing to assess and mitigate
potential limitations on further increases in muon beam intensity, and results will be
presented in a forthcoming publication.

% --------------- Acknowledgments ----------
\begin{acknowledgments}
We thank the NA64 Collaboration for the support during the test with their detector, and our colleagues contributing to Xsuite and \textsc{BDSIM}/pybdsim for support and discussions.

\end{acknowledgments}

% --------------- References ---------------
\bibliographystyle{apsrev4-2}
\bibliography{H4M2refs}

\end{document}